\documentclass[aps,prl,twocolumn,superscriptaddress,nobibnotes]{revtex4-2}
\usepackage{graphicx}
\usepackage{amssymb, amsmath}
\usepackage{booktabs}
\usepackage[usenames]{color}
\usepackage{bm}
\usepackage{multirow}
\usepackage{dcolumn}
\usepackage{hyperref}
\usepackage{enumerate}
\usepackage[mathlines]{lineno}
\usepackage[]{mdframed}
\usepackage[normalem]{ulem}
\hypersetup{
    colorlinks=true,
    linkcolor=blue,
    filecolor=gray,      
    urlcolor=blue,
    citecolor=blue,
}

\begin{document}

\title{AI-accelerated metallized $\sigma$-bonding screening for superconductor discovery}

\newcommand{\thuphy}{State Key Laboratory of Low Dimensional Quantum Physics and Department of Physics, Tsinghua University, Beijing, 100084, China}
\newcommand{\thuias}{Institute for Advanced Study, Tsinghua University, Beijing, 100084, China}
\newcommand{\fscqi}{Frontier Science Center for Quantum Information, Beijing, 100084, China}
\newcommand{\buaamat}{School of Materials Science and Engineering, Beihang University, Beijing, 100191, China}
\newcommand{\casphy}{Institute of Physics, Chinese Academy of Sciences, Beijing, 100190, China}
\newcommand{\hpstc}{Center for High Pressure Science and Technology Advanced Research, Beijing 100193, China}
\newcommand{\rmuphys}{School of Physics and Beijing Key Laboratory of Opto-electronic Functional Materials and Micro-nano Devices, Renmin University of China, Beijing, 100872, China}
\newcommand{\rmukl}{Key Laboratory of Quantum State Construction and Manipulation (Ministry of Education),
Renmin University of China, Beijing, 100872, China}
\newcommand{\szlab}{Suzhou National Laboratory, Suzhou 215123, China}
\newcommand{\contribute}{These authors contributed equally to this work.}

\affiliation{\thuphy}
\affiliation{\thuias}
\affiliation{\fscqi}
\affiliation{\buaamat}
\affiliation{\casphy}
\affiliation{\hpstc}
\affiliation{\rmuphys}
\affiliation{\rmukl}
\affiliation{\szlab}

\author{Zechen \surname{Tang}}
\thanks{\contribute}
\affiliation{\thuphy}

\author{Wen-Han \surname{Dong}}
\thanks{\contribute}
\affiliation{\thuphy}

\author{Baochun \surname{Wu}}
\thanks{\contribute}
\affiliation{\thuphy}
\affiliation{\szlab}

\author{Jian-Feng \surname{Zhang}}
\affiliation{\hpstc}

\author{Yuxiang \surname{Wang}}
\affiliation{\thuphy}
\affiliation{\thuias}

\author{Yang \surname{Li}}
\affiliation{\thuphy}

\author{Honggeng \surname{Tao}}
\affiliation{\thuphy}

\author{Qiyu \surname{Zeng}}
\affiliation{\thuphy}

\author{Chong \surname{Wang}}
\affiliation{\thuphy}
\affiliation{\fscqi}

\author{Chen \surname{Si}}
\affiliation{\buaamat}
\affiliation{\szlab}

\author{Zhong-Yi \surname{Lu}}
\affiliation{\rmuphys}
\affiliation{\rmukl}

\author{Wenhui \surname{Duan}}
\email{duanw@tsinghua.edu.cn}
\affiliation{\thuphy}
\affiliation{\thuias}
\affiliation{\fscqi}
\affiliation{\szlab}

\author{Tao \surname{Xiang}}
\email{txiang@iphy.ac.cn}
\affiliation{\casphy}

\author{Yong \surname{Xu}}
\email{yongxu@mail.tsinghua.edu.cn}
\affiliation{\thuphy}
\affiliation{\fscqi}
\affiliation{\szlab}

\begin{abstract}
The computational discovery of phonon-mediated superconductors is hindered by the prohibitive cost of density functional perturbation theory (DFPT). Here, guided by the metallized $\sigma$-bonding picture, we introduce the $\sigma$-bonding density of states ($\sigma$DOS) as an efficient physical descriptor to identify high-transition-temperature ($T_{\mathrm{c}}$) superconductors from density functional theory (DFT)-level electronic structure without explicit DFPT calculations. The evaluation of $\sigma$DOS can be further accelerated by a deep-learning DFT Hamiltonian method, enabling efficient large-scale screening for superconductors. Screening 2 million materials, we identify B$_{13}$Se as an ambient-pressure superconductor candidate with predicted $T_{\mathrm{c}} > 40$~K, together with a family of high-$T_{\mathrm{c}}$ B$_{13}X$ candidates, supporting the effectiveness of this discovery strategy. By bridging physics priors with AI acceleration, this study delivers an efficient and generalizable route for computational materials discovery in the AI era.
\end{abstract}

\maketitle

The search for superconductors with high transition temperatures ($T_\text{c}$) has long been a central topic in physics and materials science because of its importance for both fundamental science and emerging applications, including low-power electronics and quantum computing. However, discovering such materials remains highly challenging despite more than a century of research. For phonon-mediated superconductors, first-principles methods have enabled reliable computational discovery. Specifically, density functional perturbation theory (DFPT) evaluates the electron-phonon coupling (EPC)~\cite{giustino2017electron}, while subsequent $T_{\text{c}}$ predictions are enabled by the Migdal-Eliashberg theory~\cite{eliashberg1960interactions}, the Allen-Dynes approximation~\cite{allen1975transition}, or superconducting density functional theory~\cite{oliveira1988density}. The computational approach has successfully predicted numerous superconductors~\cite{dolui2024feasible}, particularly the high-$T_\text{c}$ hydride superconductors under high pressure~\cite{duan2014pressure,peng2017hydrogen,liu2017potential}. Unfortunately, DFPT calculations are prohibitively expensive, typically scaling as $O(N^4)$ with system size $N$, and are restricted to very small systems containing at most 10--20 atoms per primitive cell~\cite{lilia20222021}. This severely limits the computational discovery of high-$T_\text{c}$ superconductors.

Artificial intelligence (AI) offers new opportunities to accelerate superconductor discovery. DFPT-generated databases can be used to train surrogate AI models that predict superconductivity-related quantities, such as $T_\text{c}$~\cite{stanev2018machine,hutcheon2020predicting,wang2026computational,han2025invdesflow,han2025invdesflowAL,ouyang2025high,yao2026superconductivity} or EPC strength~\cite{tran2023machine,gibson2025accelerating,cerqueira2024sampling}, from material structures. However, the predictive performances of such models are limited by the scarce training data for known superconducting materials, while the black-box nature of direct property prediction models hinders physical interpretability and reliable extrapolation across the complex chemical space. By contrast, more physically interpretable neural-network models can be trained to learn fundamental DFPT quantities, such as the perturbation of the density functional theory (DFT) Hamiltonian induced by lattice distortions~\cite{deeph-dfpt2024}. Although this approach holds the potential for improved predictive accuracy and generalizability on unseen structures, it remains too costly for large-scale, high-throughput material screening. Thus, physics-guided, AI-accelerated approaches that are both accurate and computationally efficient remain elusive.

\begin{figure*}
\includegraphics[width=0.7\linewidth]{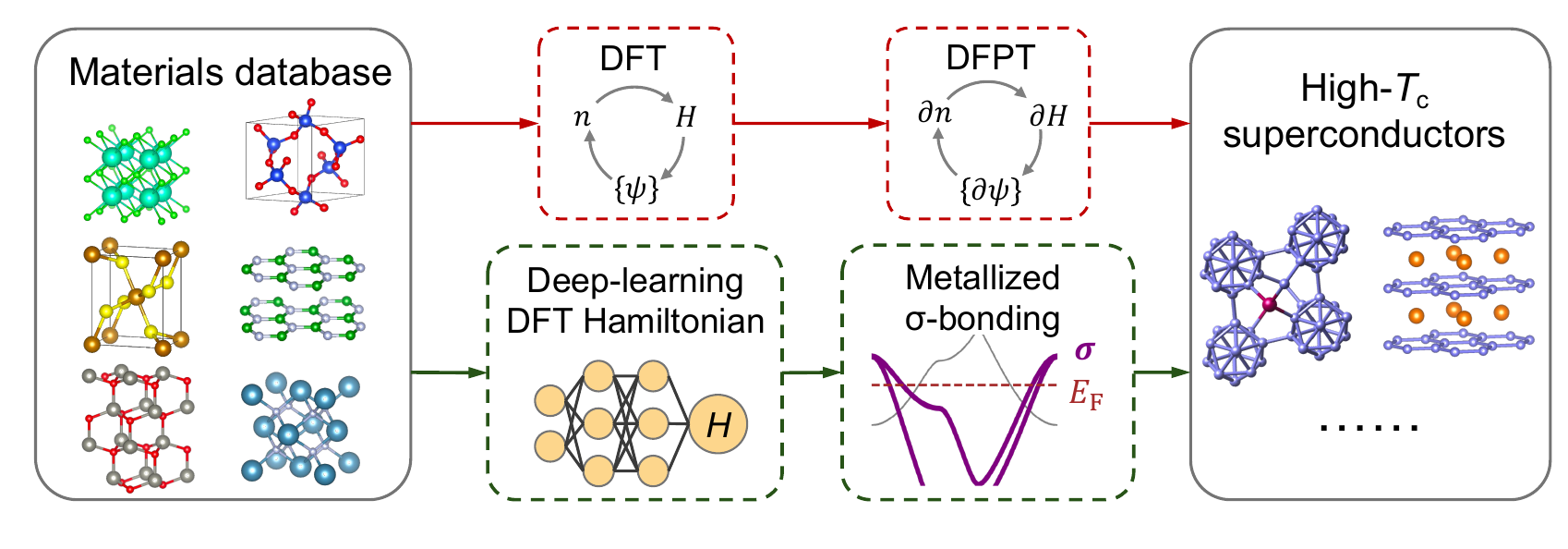}
\caption{\label{fig1}
Workflows for searching phonon-mediated superconductors. Traditional approaches rely on density functional theory (DFT) together with computationally more expensive density functional perturbation theory (DFPT). The physics-guided, AI-accelerated approach introduced in this work replaces DFPT calculations with metallized $\sigma$-bonding screening based on DFT electronic structure, and bypasses DFT calculations using a deep-learning DFT Hamiltonian approach, enabling efficient, large-scale screening for high-$T_{\mathrm{c}}$ superconductors.  
}
\end{figure*}

In this work, guided by the metallized $\sigma$-bonding picture of high-$T_\text{c}$ superconductors, we introduce a new electronic-structure descriptor, the $\sigma$-bonding density of states ($\sigma$DOS), for identifying candidate materials with strong EPC and high $T_\text{c}$ from DFT-level electronic structure, without explicit DFPT calculations. We validate the reliability of this descriptor through systematic studies of representative superconductors, and accelerate its evaluation using the deep-learning DFT Hamiltonian (DeepH) approach, further bypassing DFT simulations~\cite{deeph2022,deephe32023,deeph22024}. The screening workflow combining physics guidance and AI acceleration is demonstrated in Fig.~\ref{fig1}. By screening 2 million candidate materials, we identify B$_{13}$Se as a metallized $\sigma$-bonding-driven superconductor with predicted ambient-pressure $T_\text{c} > 40$~K, along with a family of high-$T_\text{c}$ B$_{13}X$ candidates generated by elemental substitution. Electronic structure analysis reveals clear signatures of metallized $\sigma$-bonding and unusually strong EPC in these candidate materials, further supporting the effectiveness of our physics-guided, AI-accelerated approach.

Costly DFPT calculations constitute the primary computational bottleneck in superconductor discovery. While DFPT is indispensable for {\it ab initio} prediction of $T_{\mathrm{c}}$, bypassing it during the initial screening stage is critical for high throughput exploration of large materials databases. For that, we use the metallized $\sigma$-bonding picture of high-$T_{\mathrm{c}}$ superconductors as a guiding principle, which suggests that strong superconducting pairing can arise when $\sigma$-bonding states are metallized~\cite{sigmabonding2015,mgb2pickett2001}. 
This can be rationalized by the fact that $\sigma$-bonding states typically exhibit strong EPC or antiferromagnetic fluctuations, as observed in cuprate superconductors \cite{sigmabonding2015}, which favors high $T_{\text{c}}$. The physical picture not only explains strong superconductivity in known systems, such as MgB$_2$, but also guides the design of new superconductors~\cite{LiBC2002,dopeddiamond2008,Li3B4C22015,KB2H82021,MB3C32021,B12X2025}. Nevertheless, previous applications of this picture have remained largely qualitative, lacking a quantitative framework suitable for first-principles high-throughput screening.

\begin{figure}
\includegraphics[width=1.0\linewidth]{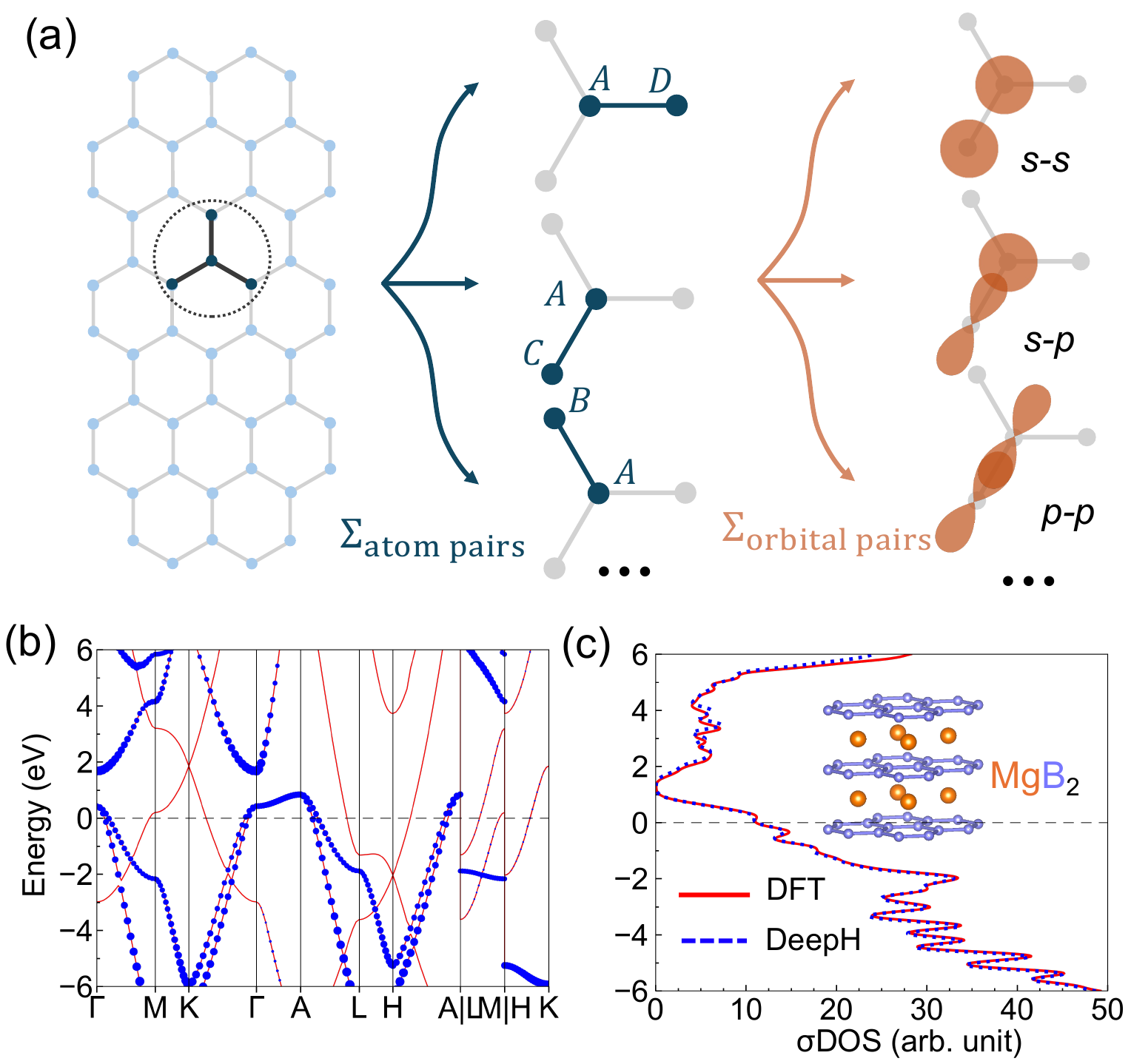}
\caption{\label{fig2} (a) $\sigma$-bonding contributions summed over neighboring atom pairs and bond-aligned orbital pairs. (b) Band structure of MgB$_2$, with state-resolved $\sigma$-bonding strength denoted by blue dots. (c) $\sigma$-bonding density of states ($\sigma$DOS) of MgB$_2$ obtained from DFT and predicted by the deep-learning DFT Hamiltonian (DeepH) method.}
\end{figure}

We first introduce a quantitative metric to characterize metallized $\sigma$ bonding, subject to the following requirements: (i) energy specificity, as only electronic states near the Fermi level contribute to superconductivity; (ii) bond specificity, as $\sigma$-bonding occurs only between neighboring atom pairs; and (iii) orbital specificity, as only bond-aligned orbital pairs contribute to $\sigma$ bonding. Thus, a physical descriptor, the $\sigma$-bonding DOS ($\sigma$DOS), is proposed to measure the strength of metallized $\sigma$ bonding:
\begin{align}
\label{sigma_bonding_dos}
\sigma\text{DOS}(E)=\frac{N_{\text{atom}}}{\Omega_{\text{BZ}}}\int \mathrm{d}\mathbf{k}\sum_n \sigma_{n\mathbf{k}}\delta(E-\epsilon_{n\mathbf{k}}),
\end{align}
where $\epsilon_{n\mathbf{k}}$ and $\sigma_{n\mathbf{k}}$ denote the eigenvalue and state-resolved $\sigma$-bonding strength, respectively, of the $n$-th electronic state at momentum $\mathbf{k}$, while $N_{\text{atom}}$ and $\Omega_{\text{BZ}}$ represent the number of atoms per unit cell and the Brillouin zone volume. Since the $\sigma$-bonding analysis requires real-space projection, basis sets using localized atomic-like orbitals are suitable for the purpose. The approach can nevertheless be generalized to other basis sets, such as plane waves, by projecting DFT Hamiltonians to atomic orbitals~\cite{deeph-pw}. Thus, Bloch states are  expressed as linear combinations of atomic-like orbitals $\{\phi_{I\alpha}\}$, where $I$ and $\alpha$ denote atom and orbital indices, respectively, with $c^{I\alpha}_{n\mathbf{k}}$ as the corresponding expansion coefficients. The key quantity $\sigma_{n\mathbf{k}}$ is obtained by summing over atom and orbital pairs $(I\alpha, J\beta)$, as illustrated in Fig.~\ref{fig2}(a):
\begin{align}
\label{sigma_bonding_state}
\sigma_{n\mathbf{k}}=\sum_{IJ}\sum_{\alpha\beta}\left|c^{I\alpha}_{n\mathbf{k}}\right|^2\left|c^{J\beta}_{n\mathbf{k}}\right|^2\rho_{IJ}\gamma_{\alpha\beta},
\end{align}
where the requirements of bond specificity and orbital specificity are enforced through the factors $\rho_{IJ}$ and $\gamma_{\alpha\beta}$, respectively. $\rho_{IJ}$ is nonzero only for neighboring atom pairs involving specific chemical elements. $\gamma_{\alpha\beta}$ ensures orbital compatibility by transforming $c^{I\alpha}_{n\mathbf{k}}$ and $c^{J\beta}_{n\mathbf{k}}$ to a local coordinate system with $z$-axis aligned with the bonding direction; only symmetry-compatible orbitals (e.g., $s, p_z, d_{z^2}$) contribute. The method details are provided in the Supplemental Section 1~\cite{supp}.

While the definition of $\sigma$DOS is not unique, its application as a quantitative metric in high-throughput screening requires invariance under translation, rotation, and cell-size variations, thereby enabling consistent evaluation across diverse crystal structures. The fact that $\{\phi_{I\alpha}\}$ possess spherical harmonic angular parts facilitates transformation of $c_{n\mathbf{k}}^{I\alpha}$ to local coordinates. This enables $\gamma_{\alpha\beta}$ to be evaluated in local coordinate systems, and also ensures invariance under structural translation and rotation. Moreover, the normalization factor $\frac{N_{\text{atom}}}{\Omega_{\text{BZ}}}$ in Eq.~\ref{sigma_bonding_dos} ensures invariance with respect to the choice of unit cell, enabling consistent comparisons across structures with different unit-cell sizes (see Supplemental Section~1~\cite{supp}).

While $\sigma$DOS serves as an overall quantitative measure of metallized $\sigma$-bonding strength, the 
state-resolved quantity $\sigma_{n\mathbf{k}}$ is also useful for analysis. We demonstrate this for MgB$_2$ (Fig.~\ref{fig2}(b)), a well-known superconductor with metallized $\sigma$-bonding, in which boron atoms form layered graphene-like sheets. The hole-doped $\sigma$-bands induce enhanced superconductivity with $T_{\text{c}}=39$~K~\cite{mgb2pickett2001}. Consistent with this physics, our framework clearly distinguishes the $\sigma$ and $\pi$-bands, capturing the metallized $\sigma$-bonding characteristics. MgB$_2$ exhibits a high $\sigma$DOS of 10.6 at $E=E_F$, a threshold exceeded by very few materials even in established computational superconductor database (Fig.~S1(h)~\cite{supp}), highlighting the discriminative power of this metric. The framework's robustness extends to other metallized $\sigma$-bonding-driven superconductors, as detailed in Supplemental Section~2~\cite{supp}. Furthermore, benchmarking against the computational $T_{\mathrm{c}}$ database from Ref.~\cite{cerqueira2024sampling} reveals a strong positive correlation between $T_{\mathrm{c}}$ and $\sigma$DOS($E_F$). Specifically, 418 of 570 candidates with $\sigma$DOS($E_F$) $> 10$ exhibit predicted $T_{\mathrm{c}} > 10$~K, validating the $\sigma$DOS metric as a reliable indicator for high-$T_{\text{c}}$ superconductivity. This trend is visualized in the $T_{\mathrm{c}}$--$\sigma$DOS($E_F$) distribution and quantified via a sliding-window average (Fig.~S1(g,h)~\cite{supp}). Notably, the rarity of materials with large $\sigma$DOS($E_F$) underscores the metric's utility as an efficient pre-screening filter, effectively narrowing the candidate pool (Fig.~S1(h)~\cite{supp}).

Although $\sigma$DOS enables screening based on DFT-level electronic structure, application to millions of materials remains costly. To bypass explicit DFT simulations for further speedup, we employ the deep-learning DFT Hamiltonian method (DeepH)~\cite{deeph2022,deephe32023,xDeepH2023}. DeepH models the DFT Hamiltonian as a function of material structure using a neural network, enabling efficient retrieval of wavefunctions and properties like $\sigma$DOS via postprocessing~\cite{tang2025deepreview}. Recently, DeepH has evolved to handle ``universal" database structures across vast chemical elements and crystal structures~\cite{2024-deeph-umm}. We illustratively present the accuracy and generalization capability of the universal DeepH model using $\sigma$DOS for MgB$_2$ (Fig.~\ref{fig2}(c)). DeepH-derived $\sigma$DOS matches DFT nearly exactly, underscoring model fidelity. DeepH's predictive capability for $\sigma$DOS is further validated on three additional proposed superconductors (Fig.~S1~\cite{supp}), all absent from the training set (Materials Project~\cite{jain2013commentary}), demonstrating its transferability to unseen structures.

The universal DeepH model is applied to screen 2 million structures from the Alexandria and GNoME databases~\cite{alexandria2023,gnome2023}. Candidates with $\sigma\text{DOS}(E_{\text{F}}) > 10$ are validated using DFPT, yielding 31 promising superconductors ($T_\text{c}$ ranges from 10 to 140~K). The high-throughput discovery, along with the technical details and accuracy benchmarks of the universal DeepH model, is reported in a companion study~\cite{2026-deeph-umm}. Among these, the previously unexplored boride B$_{13}$Se emerges as a notable case, with a large $\sigma\text{DOS}(E_{\text{F}})$ of 67.2 (Fig.~S2~\cite{supp}). Hereafter, we focus on B$_{13}$Se as a representative metallized $\sigma$-bonding-driven superconductor, employing refined computational parameters beyond the high-throughput screening.

\begin{figure*}
\includegraphics[width=0.8\linewidth]{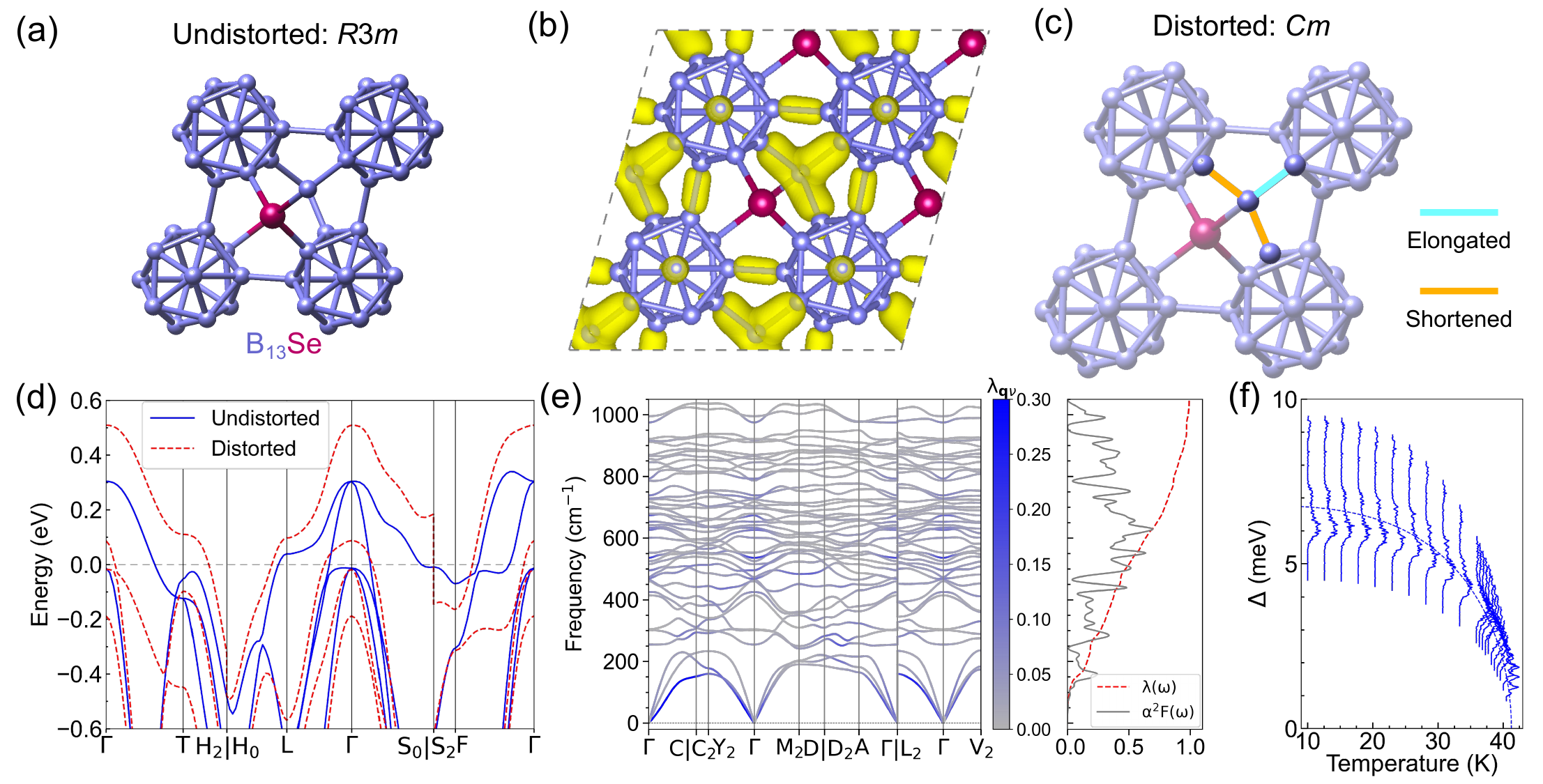}
\caption{\label{fig3} (a,c) Atomic structures and (d) band structures of undistorted and distorted B$_{13}$Se, which consists of one B$_{12}$ icosahedron and two interstitial atoms (B and Se) per unit cell. Minor displacements of the interstitial boron lower the space-group symmetry from $R3m$ to $Cm$, resulting in pronounced band splitting. (b) local density of states of undistorted B$_{13}$Se near the Fermi level.  (e) Phonon dispersion of distorted B$_{13}$Se with state-resolved electron-phonon coupling (EPC) strength ($\lambda_{\textbf{q}\nu}$) (left panel), and Eliashberg spectral function $\alpha^2F(\omega)$ together with cumulative EPC strength $\lambda(\omega)$ (right panel). (f) Temperature-dependent superconducting gap $\Delta$ obtained from anisotropic Migdal-Eliashberg theory.}
\end{figure*}

As shown in Fig.~\ref{fig3}(a), the undistorted structure of B$_{13}$Se adopts a trigonal space group $R3m$ (No. 160) with lattice parameters $a=5.226$ \AA\ and $\alpha=68.8^\circ$. Its structure consists of a B$_{12}$-icosahedral framework, one interstitial B atom and another interstitial Se atom. In each unit cell, 45 valence electrons occupy the lowest 23 bands, yielding an intrinsic metallic band structure (Fig.~\ref{fig3}(d), Fig.~S8(a)~\cite{supp}). The electronic states near the Fermi level are dominated by $\sigma$-bondings involving both B$_{12}$-B$_{12}$ and B$_{12}$-interstitial B connections, as evidenced by the local density of states near the Fermi level (Fig.~\ref{fig3}(b)) and the $\sigma$-bonding-like Wannier functions (Fig.~S3~\cite{supp}). The DFT-calculated and DeepH-predicted band structures and $\sigma$DOS of B$_{13}$Se are shown in Fig.~S2~\cite{supp}. The close agreement between DFT and DeepH validates DeepH’s predictive accuracy and effectiveness in accelerating superconductor screening.

While a large DOS at the Fermi level indicates strong EPC, it may also trigger instabilities. We therefore assess the stability of B$_{13}$Se against both magnetic states and structural distortions. Magnetism is ruled out by explicit verification calculations of ground state (Fig.~S4~\cite{supp}). Structurally, we identify a slightly distorted phase in space group $Cm$, characterized by symmetry breaking of the bonds between B$_{12}$ and interstitial boron: two bonds shorten by 0.02~\AA\ and one elongated by 0.05~\AA\ (Fig.~\ref{fig3}(c)). This distortion lifts the $C_3$-protected band degeneracy (Fig.~\ref{fig3}(d)), lowering the electronic occupation energy and stabilizing the structure by 0.8~meV/atom (Supplemental Section~4~\cite{supp}). We thus select the distorted $Cm$ phase for subsequent analysis.

The superconducting properties are investigated through DFPT, with computational details presented in Supplemental Section~8~\cite{supp}. As shown in the left panel of Fig.~\ref{fig3}(e), the phonon spectrum of distorted B$_{13}$Se exhibits no imaginary modes, confirming its dynamical stability. Further calculations of the Eliashberg spectral function $\alpha^2F(\omega)$ and cumulative EPC strength $\lambda(\omega)$ (Fig.~\ref{fig3}(e) right) reveal a stronger EPC strength $\lambda = 0.99$ than the well-known superconducting boride MgB$_{2}$ ($\lambda = 0.78$~\cite{margine2013anisotropic}). $\lambda(\omega)$ increases smoothly with $\omega$, indicating broad contributions from multiple phonon branches, as reflected by the state-resolved $\lambda_{\mathbf{q}\nu}$ (Fig.~\ref{fig3}(e)). We evaluate $T_{\mathrm{c}}$ using anisotropic Eliashberg theory \cite{eliashberg1960interactions} with an effective Coulomb repulsion $\mu^*=0.13$ (a typical value in superconductivity calculations~\cite{bercx2025charting}). This predicts single-gap superconductivity with $T_{\mathrm{c}} = 41$ K for B$_{13}$Se (Fig.~\ref{fig3}(f)). The $T_{\mathrm{c}}$ dependence on $\mu^*$ is discussed in Fig.~S9~\cite{supp}.

\begin{figure}
\includegraphics[width=1.0\linewidth]{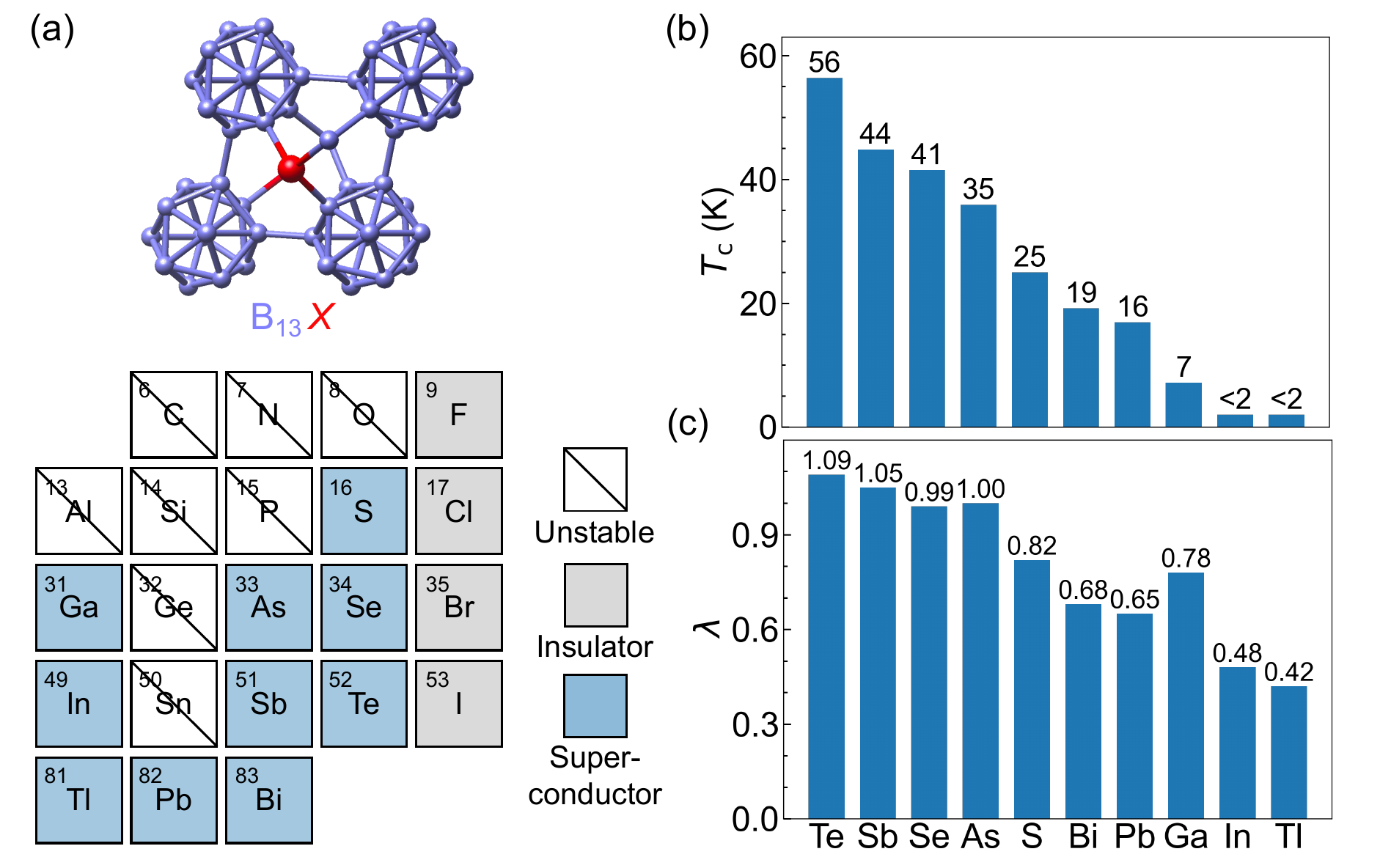}
\caption{\label{fig4} (a) Superconducting properties of the B$_{13}X$-family materials, where $X$ denotes a non-radioactive $p$-block element. (b, c) Predicted superconducting transition temperature ($T_{\mathrm{c}}$) and electron-phonon coupling strength ($\lambda$) of B$_{13}X$ for varying element $X$.}
\end{figure}

The prediction of B$_{13}$Se indicates the feasibility of high-$T_{\text{c}}$ superconductor discovery with the $\sigma$DOS indicator, and motivates the exploration of a class of metallized $\sigma$-bonding-driven superconductors, the B$_{13}X$ family. Inspired by the fourfold covalent coordination of the interstitial Se atom in B$_{13}$Se, we investigate B$_{13}X$ compounds with $X$ being a non-radioactive $p$-block element. As shown in Fig.~\ref{fig4}(a), ten superconducting candidates are identified, five of which exhibit predicted critical temperatures $T_{\mathrm{c}} > 20$~K within anisotropic Migdal-Eliashberg theory using $\mu^* = 0.13$ (Fig.~\ref{fig4}(b)), accompanied with strong EPC (Fig.~\ref{fig4}(c)). For $X$ = Se, As, and S, a lower-energy $Cm$-distorted phase is found, whereas the remaining compounds adopt the $R3m$ space group. The band structures and phonon spectra of the high-$T_{\mathrm{c}}$ candidates (B$_{13}X$, $X=$S, As, Se, Sb and Te) are summarized in Fig.~S8~\cite{supp}, together with an analysis of the $T_{\mathrm{c}}$ dependence on the empirical parameter $\mu^*$ in Fig.~S9~\cite{supp}. Finally, the thermodynamic stability of higher-$T_{\text{c}}$ candidates including B$_{13}X$, $X=$S, As, Se, Sb and Te, is investigated, indicating favorable synthesizability of B$_{13}$Se and B$_{13}$S (Supplemental Section~5~\cite{supp}).

In summary, we propose a physics-guided, AI-accelerated method for superconductor discovery based on the metallized $\sigma$-bonding picture. Using this approach, we discover the B$_{13}X$-family materials as a prototype class of high-$T_\text{c}$ phonon-mediated superconductors. Looking ahead, deep-learning approaches can be further improved, for example through the use of advanced hybrid functionals to enhance accuracy~\cite{deeph-hybrid} or DeepH-DFPT~\cite{deeph-dfpt2024} to accelerate the final validation calculations. Moreover, the $\sigma$DOS descriptor can itself serve as a useful machine-learning target, and the greater availability of training data may facilitate the development of generative AI models for discovering high-$T_\text{c}$ superconductors. More broadly, the physics-guided, AI-accelerated strategy could be extended to other material properties, broadening the scope of physics- and AI-driven materials discovery.

\begin{acknowledgments} 
This work was supported by the Basic Science Center Project of NSFC (Grant No. 52388201), the National Key Basic Research and Development Program of China (Grant No. 2024YFA1409100), the Beijing Municipal science and Technology Commission, Administrative Commission of Zhongguancun Science Park (grant no. Z251100003625025), the Fundamental and Interdisciplinary Disciplines Breakthrough Plan of the Ministry of Education of China (Grant No. JYB2025XDXM408), the National Natural Science Foundation of China (Grants No. 12334003, No. 12421004, No. 12488201, No. 124B2072, No. 12504078, and No. 12361141826), the National Key Basic Research and Development Program of China (Grant No. 2023YFA1406400), Quantum Science and Technology-National Science and Technology Major Project (Grants No. 2023ZD0300500 and No. 2021ZD0301800), the China Postdoctoral Science Foundation (Grant no. 2025M773367), ShanghaiTech AI4S Initiative SHTAI4S202504, Beijing Key Laboratory of Quantum AI, and the Beijing Advanced Innovation Center for Future Chip (ICFC). The calculations were performed at National Supercomputer Center in Tianjin using the Tianhe new generation supercomputer.

\end{acknowledgments}

\end{document}